%% file: pid-v2-62.19.tex
\documentclass[twocolumn,showpacs,aps,superscriptaddress]{revtex4}

\usepackage{graphicx}
\usepackage{dcolumn}
\usepackage{amsmath}
\usepackage{supertabular}
\usepackage{multirow}
\usepackage{multicols}

\begin{document}

\title{ Mass, quark-number, and $\sqrt{s_{_{NN}}}$ dependence of the
  second and fourth flow harmonics in ultra-relativistic
  nucleus-nucleus collisions }

\include{authors}

\date{\today}

\begin{abstract} %
We present STAR measurements of the azimuthal anisotropy parameter
$v_2$ for pions, kaons, protons, $\Lambda$, $\overline{\Lambda}$,
$\Xi+\overline{\Xi}$, and $\Omega + \overline{\Omega}$, along with
$v_4$ for pions, kaons, protons, and $\Lambda + \overline{\Lambda}$ at
mid-rapidity for Au+Au collisions at $\sqrt{s_{_{NN}}}=62.4$ and
200~GeV. The $v_2(p_T)$ values for all hadron species at 62.4 GeV are
similar to those observed in 130 and 200 GeV collisions. For observed
kinematic ranges, $v_2$ values at 62.4, 130, and 200 GeV are as little
as 10\%--15\% larger than those in Pb+Pb collisions at
$\sqrt{s_{_{NN}}}=17.3$ GeV. At intermediate transverse momentum
($p_T$ from 1.5--5~GeV/c), the 62.4~GeV $v_2(p_T)$ and $v_4(p_T)$
values are consistent with the quark-number scaling first observed at
200 GeV. A four-particle cumulant analysis is used to assess the
non-flow contributions to pions and protons and some indications are
found for a smaller non-flow contribution to protons than pions.
Baryon $v_2$ is larger than anti-baryon $v_2$ at 62.4 and 200 GeV
perhaps indicating either that the initial spatial net-baryon
distribution is anisotropic, that the mechanism leading to transport
of baryon number from beam- to mid-rapidity enhances $v_2$, or that
anti-baryon and baryon annihilation is larger in the in-plane
direction.

\end{abstract}


\maketitle

\vspace{0.5cm}

\section{Introduction}

In non-central heavy-ion collisions, the overlapping area has a
long axis and a short axis. Re-scattering amongst the system's
constituents converts the initial coordinate-space anisotropy to a
momentum-space anisotropy~\cite{OllitraultSorge, firstv2,rhicv2}. The
\textit{spatial} anisotropy decreases as the evolution progresses so
that the \textit{momentum} anisotropy is most sensitive to the early
phase of the evolution --- before the spatial asymmetry is
washed-out~\cite{Kolb:2000sd}.

Ultra-relativistic Au+Au collisions at Brookhaven National
Laboratory's Relativistic Heavy Ion Collider (RHIC)~\cite{RHIC} are
studied in part to deduce whether quarks and gluons become deconfined
during the early, high energy-density phase of these collisions. Since
the azimuthal momentum-space anisotropy of particle production is
sensitive to the early phase of the collision's evolution, observables
measuring this anisotropy are especially interesting. The azimuth
angle ($\phi$) dependence of particle momentum distributions can be
expressed in the form of a Fourier series: $dN/d\phi \propto 1 +
\sideset{}{_n}\sum\nolimits 2v_n\cos n\left (\phi-\Psi_{RP} \right )$,
where $\Psi_{RP}$ is the reaction-plane
angle~\cite{note,Voloshin:1994mz}. The Fourier coefficients $v_n$ can
be measured and used to characterize the azimuthal anisotropy of
particle production.

Measurements at two higher RHIC energies ($\sqrt{s_{_{NN}}}=130$ and
200~GeV) established that charged hadron $v_2$ rises with $p_T$ for
$p_T<2$~GeV/c and then saturates~\cite{Adler:2002pu,Adler:2002ct}. As
predicted by the hydrodynamic calculations~\cite{hydroPasi01,
  hydroShuryak01} --- where local thermal equilibrium is assumed ---
$v_2$ at low $p_T$ ($p_T < 1$~GeV/c) shows a characteristic dependence
on particle mass~\cite{PIDv2,Adams:2004bi}.  The $v_2$ values at
$\sqrt{s_{_{NN}}}=130$ and 200 GeV are as large as those predicted by
hydrodynamic calculations. The $v_2$ values measured at
$\sqrt{s_{_{NN}}}=17.3$~GeV~\cite{NA49}, the top energy of the Super
Proton Synchrotron (SPS) at CERN, however, are below the hydrodynamic
models predictions. In this paper, we compare $v_2$ at
$\sqrt{s_{_{NN}}}=17.3$ and 200 GeV to new measurements at
$\sqrt{s_{_{NN}}}=62.4$~GeV that provide a link between the top RHIC
energy and the top SPS energy.

In 200 GeV collisions, kaon, proton, $\Lambda+\overline{\Lambda}$ and
$\Xi + \overline{\Xi}$ $v_2(p_T)$ at intermediate $p_T$ depends on the
number $n_q$ of constituent quarks in the corresponding
hadron~\cite{scalingv2}. A scaling law --- motivated by
constituent-quark coalescence or recombination models --- can account
for the observed splitting between baryons and mesons for $v_2$ in
this intermediate region~\cite{scalingv2,CoalVoloshinv2}. Within these
models, hadron $v_2$ ($v_2^h$) is related to the $v_2$ of quarks
($v_2^q$) in a quark-gluon phase by the relationship: $v_2^h(p_T^h)
\approx n_q v_2^q(n_q p_T^q)$~\cite{reco}. Intermediate
$p_T$ baryon yields also increase with collision centrality more
rapidly than meson yields~\cite{scalingv2, Adler:2003kg}: a behavior
also expected from coalescence or recombination
models~\cite{reco}. These models suggest that the large $v_2$ values
at intermediate $p_T$ are developed during a pre-hadronic phase --- a
conclusion supported by the recent discovery that multi-strange
baryons, thought to have smaller hadronic
cross-sections~\cite{Biagi:1980ar,msb130}, attain $v_2$ values
apparently as large as protons or hyperons~\cite{msbv2}. Measurements
of $v_2$ for identified particles may, therefore, help reveal whether
$v_2$ is developed in a deconfined quark-gluon phase and can test
whether these possible conclusions may still be valid at lower
$\sqrt{s_{_{NN}}}$.

This article is organized as follows: in Sec.~\ref{ex} we briefly
describe the STAR detector.  The analysis procedures are presented in
Sec.~\ref{an}.  In Sec.~\ref{re} we present our results. This section
includes subsections discussing systematic uncertainties, baryon
versus anti-baryon $v_2$, quark-number scaling, the fourth harmonic
$v_4$, and the collision energy dependence of $v_2$. Our conclusions
are then presented in Sec.~\ref{co}.

\section{Experiment}
\label{ex}

Our data were collected from Au~+~Au collisions at $\sqrt{s_{_{NN}}} =
62.4$ and 200~GeV with the STAR detector~\cite{STAR}. STAR's main time
projection chamber (TPC)~\cite{tpc} was used for particle tracking and
identification with supplementary particle identification provided by
time-of-flight detectors (TOF)~\cite{tof}. We analyzed events from a
centrality interval corresponding to 0\%--80\% of the hadronic
interaction cross-section. As in previous STAR
analyses~\cite{scalingv2}, we define the centrality of an event from
the number of charged tracks in the TPC having pseudo-rapidity $|\eta|
< 0.5$, $p_T>0.2$~GeV/c, a distance of closest approach to the primary
vertex (DCA) less than 2 cm, and more than 10 measured space
points~\cite{mult}. Only events with primary vertices within 30 cm of
the TPC center in the beam direction were analyzed.

STAR's main TPC covers the approximate pseudo-rapidity region $|\eta|
< 1.2$ (for collisions at its center) and $2\pi$ in azimuth angle. A
0.5 Tesla magnetic field allows charged particle $p_T$ to be measured
above 0.1 GeV/c. At the time of data taking the TOF detectors covered
$-1<\eta < 0$ and $\pi/15$ in azimuth angle. Their timing resolutions
are $\sim 110$~ps so that pions and kaons can be distinguished for
$p_T <$ 1.8~GeV/c and protons can be identified up to $p_T=3.0$~GeV/c.

\section{Analysis}
\label{an}

We identify particles using three different methods: measurement of
specific ionization-energy-loss per unit length in the TPC gas
($dE/dx$), time-of-flight measurements, and weak-decay vertex
finding. $dE/dx$ measurements for a particle with a given momentum are
used for identification at low $p_T$ and in the relativistic-rise
region ($p_T > 2.0$~GeV/c) where $dE/dx$ increases logarithmically
with $\beta \gamma$ (see Ref.~\cite{Shao:2005iu} and Fig. 26 in
Ref.~\cite{hb}). The pion sample in the relativistic-rise region is
selected based on the deviation between the measured $dE/dx$ of each
track and the expected $dE/dx$ for a pion in units of Gaussian
standard deviations ($n\sigma_{\pi}$).  For $p_T>2.0$~GeV/c, pions are
selected with $n\sigma_{\pi}>0$ (the top half of the distribution). In
this case the purity is estimated to be $98$\%.

The $v_2$ of protons is measured in this region by fitting the $dE/dx$
distribution with peaks centered at the predicted $dE/dx$ values. From
these fits we can derive the relative fractions of pions ($f_{\pi}$),
kaons ($f_{K}$) and protons ($f_{p}$) as a function of $dE/dx$. We
then measure $v_2$ for all tracks and plot it versus the $dE/dx$ of
the track. Once the relative fractions of each particle are known for
each value of $dE/dx$, and $v_2$ is know as a function of $dE/dx$,
$v_{2}(dE/dx)$ can be fit with function:
\begin{equation}
v_{2}(dE/dx) = f_{\pi}v_{2,\pi} + f_{K}v_{2,K} + f_{p}v_{2,p}, 
\end{equation}
where the $v_2$ values for each species ($v_{2,\pi}$, $v_{2,K}$, and
$v_{2,p}$) are parameters in the fit and $f_{\pi}$, $f_K$, and $f_p$
which are extracted from the $dE/dx$ distribution, are part of the fit
function. In the relativistic rise region, kaons do not dominate the
$dE/dx$ distribution for any value of $dE/dx$, so their $v_2$ values
are poorly constrained and are not presented here. We estimated the
systematic error on the proton $v_2$ by varying the relative fractions
of the different particles within reasonable limits. The relative
change in the proton $v_2$ ($\delta v_2/v_2$) was less than 3\%. The
shape and width of the peaks are determined from samples of particles
identified by other means, \textit{e.g.} TOF measurements and $K_S^0$
or $\Lambda$ decay daughters.

The reaction-plane direction is estimated for each event from the
azimuthal distribution of charged tracks. We select tracks using
criteria similar to those in Ref.~\cite{firstv2}. To avoid
self-correlations, we subtract the contribution of a given particle
from the total reaction-plane vector. For particles identified through
their decays, we subtract the contributions of all the decay
products. The reaction-plane resolution is estimated using the
sub-event method~\cite{art} and we correct the observed $v_2$ to
account for the dilution caused by imperfect resolution. The
resolution depends on the number of tracks used in the calculation and
the magnitude of $v_2$, and therefore depends on centrality. The
resolution for $\sqrt{s_{_{NN}}} = 62.4$~GeV collisions is reduced
relative to $\sqrt{s_{_{NN}}} = 200$~GeV collisions by $\approx 30$\%.
For 62.4 GeV Au+Au collisions it reaches a maximum value of
approximately 0.73 in the 10\%--40\% centrality interval.

\section{Results}
\label{re}

\begin{figure*}[htbp]
\centering\mbox{
\includegraphics[width=0.75\textwidth]{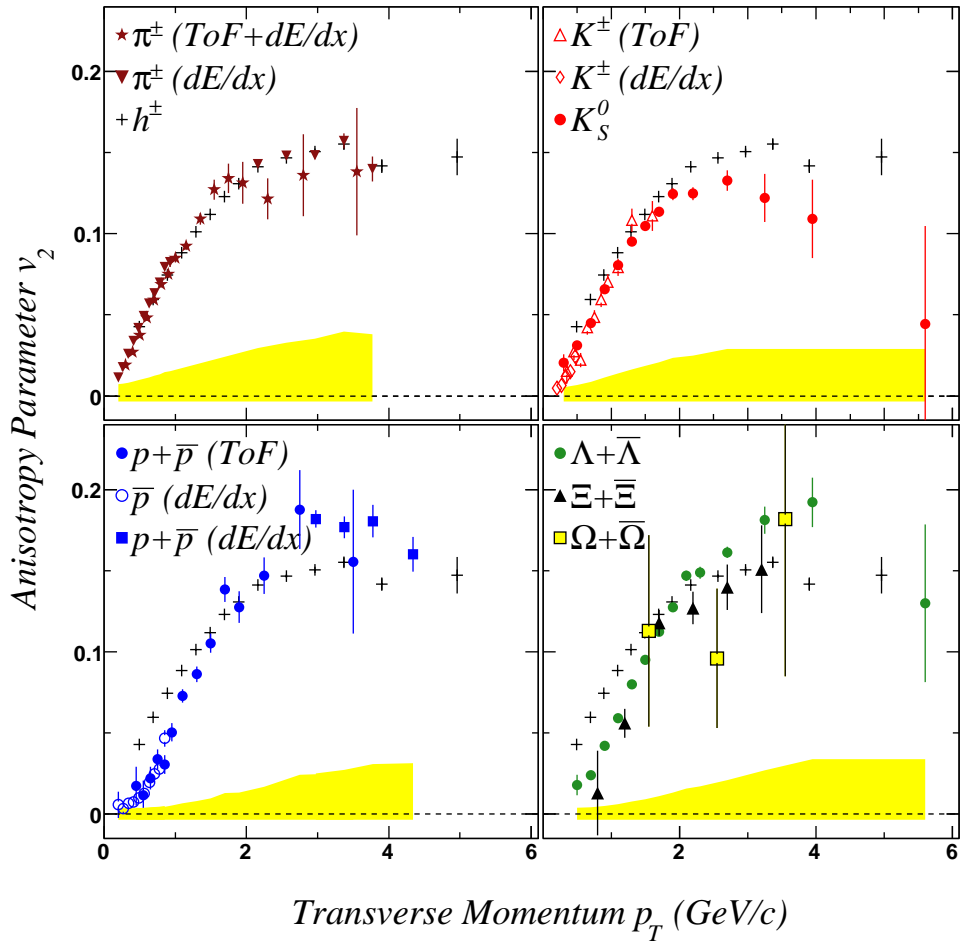}}
\caption{ (color online). Minimum-bias (0--80\% of the collision cross
  section) $v_{2}(p_T)$ for identified hadrons at $|\eta|<1$ from
  Au+Au collisions at $\sqrt{s_{_{NN}}}=62.4$~GeV. To facilitate
  comparisons between panels, $v_2$ values for inclusive charged
  hadrons are displayed in each panel. The error bars on the data
  points represent statistical uncertainties. Systematic uncertainties
  for the identified particles are shown as shaded bands around
  $v_2=0$.} \label{fig1}
\end{figure*}

In Fig.~\ref{fig1} the minimum-bias, mid-rapidity $v_2$ values are
shown for inclusive charged hadrons, pions, kaons, protons,
$\Lambda+\overline{\Lambda}$, $\Xi+\overline{\Xi}$, and
$\Omega+\overline{\Omega}$. The gross features of $v_2$ at
$\sqrt{s_{_{NN}}}=62.4$~GeV are similar to those observed at
$\sqrt{s_{_{NN}}}=200$~GeV~\cite{PIDv2,scalingv2}. For
$p_T<1.5$~GeV/c, a mass hierarchy is observed with $v_2$ smaller for
heavier particles. The $p_T$ and mass dependencies are
\textit{qualitatively} (not necessarily quantitatively) consistent
with expectations from hydrodynamic calculations that assume the
mean-free-path between interactions is zero~\cite{hydroPasi01}. For
$p_T>2$~GeV/c, $v_2$ reaches a maximum, the mass ordering is broken,
and $v_2$ for protons and hyperons tend to be larger than for either
pions or kaons. The $v_2$ values for protons and
$\Lambda+\overline{\Lambda}$ above $p_T = 2$~GeV/c are similar. In
this region, the multi-strange baryons also exhibit $v_2$ values
similar to protons.  While hadrons containing strange quarks are
expected to be less sensitive to the hadronic stage, we do not see a
statistically significant reduction in the $v_2$ values of strange
baryons compared to protons. Statistical uncertainties, however, still
do not exclude the possibility of some strangeness content dependence
for $v_2$. If $v_2$ or its hadron species dependence is developed
through hadronic interactions, $v_2$ should depend on the
cross-sections of the interacting hadrons (with hadrons with smaller
cross-sections developing less anisotropy). The large $v_2$ values for
$\Xi+\overline{\Xi}$ and $\Omega+\overline{\Omega}$ are consistent
with $v_2$ having been developed before hadronization.

\begin{figure*}[hbtp]
\centering\mbox{
\includegraphics[width=0.65\textwidth]{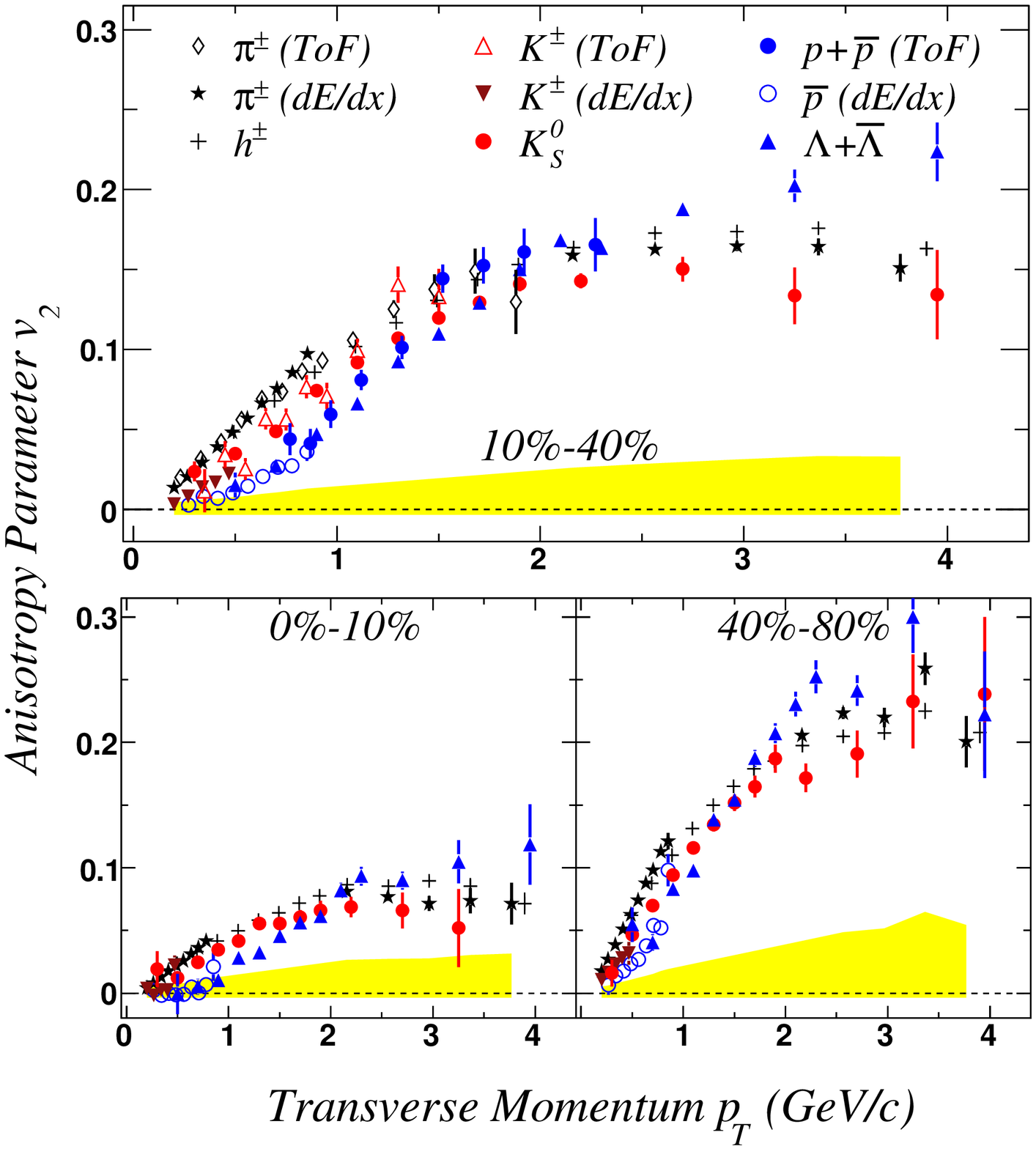}
}
\caption{ (color online). The unidentified charged hadron, charged
  pion, $K_{S}^{0}$, charged kaon, proton and
  $\Lambda+\overline{\Lambda}$ $v_2$ as a function of $p_T$ for
  10\%--40\%, 0\%--10\% and 40\%--80\% of the Au+Au interaction cross
  section at $\sqrt{s_{_{NN}}}=62.4$~GeV. Weak-decay feed-down errors
  are included in the error bars on the data points while non-flow and
  tracking error uncertainties are plotted as bands around $v_2=0$,
  which apply to all identified particles. The errors are asymmetric
  and the portion of the error band above (below) zero represents the
  negative (positive) error.} \label{fig2}
\end{figure*}

The centrality dependence of identified hadron $v_2(p_T)$ for
$\sqrt{s_{_{NN}}}=62.4$~GeV is shown in Fig.~\ref{fig2}. Similar $p_T$
and mass dependencies are observed for each of the centrality
intervals: 0\%--10\%, 10\%--40\%, and 40\%--80\%. The data from the
0\%--10\% interval are most affected by non-flow
effects~\cite{Adler:2002pu} while the 10\%--40\% interval is least
affected by these uncertainties. The particle-type dependence of
non-flow will be discussed in the following section.

\subsection{Systematic Uncertainties} 

Systematic uncertainties are shown in Fig.~\ref{fig1} as bands around
$v_2=0$. The errors are asymmetric. The portions of the band above
zero represent the negative errors so that the difference between the
measurement and zero is more visually evident. These uncertainties
take into account effects from weak-decay feed-down, tracking
artifacts, detector artifacts, and non-flow effects. Non-flow effects
are dominant. In Fig.~\ref{fig2} the tracking and non-flow systematic
uncertainties are shown as bands around $v_2=0$ and the weak-decay
feed-down uncertainties are included in the error bars on the pion
data points.

The number of tracks coming from weak-decays that are included in the
$v_2$ analysis depends on the experimental setup and track selection
criteria. Pions produced in $K_S^0$, $\Lambda$, or
$\overline{\Lambda}$ decays tend to be distributed at low $p_T$ with
$v_2$ values larger than the pions from other sources.  We have
calculated their effect on the observed pion $v_2$. We assume
exponential $m_T$ spectra for $K^0_S$ and $\Lambda$ with inverse slope
parameters of 285 and 300 MeV respectively. For relative abundances,
we take $K_S^0/(\pi^++\pi^-)$ and
$(\Lambda+\overline{\Lambda})/(\pi^++\pi^-)$ ratios of 0.06 and 0.054
respectively. The $v_2$ of $K^0_S$ and $\Lambda$ are taken from
data. We then use a full detector simulation to estimate what fraction
of the weak-decay products will fall within our detector acceptance
and pass our track selection criteria. We find that for our analysis,
feed-down will increase $v_2$ by approximately 13\% (as a fraction of
the original $v_2$) at $p_T=0.15$~GeV/c. The increase falls to
approximately 3\% relative at $p_T=0.25$~GeV/c and is negligible for
$p_T > 0.4$~GeV/c. Modifications to the observed proton $v_2$ from
$\Lambda$ and $\overline{\Lambda}$ decays are neglegible due to the
similarity of proton and hyperon $v_2$.

$v_2$ measurements can also be distorted by anti-correlations that
arise from tracking errors (\textit{e.g.} track-merging and
hit-sharing). These anti-correlations can be eliminated by correlating
tracks with $\eta>0$ $(\eta<0)$ with an event plane determined from
tracks at $\eta<-0.15$ $(\eta>0.15)$ ($\eta$-subevents). This method
also has a different sensitivity to the spurious correlations arising
from jets and resonance decays (non-flow effects discussed in the next
paragraph). In this paper, $\eta$-subevents are used to analyze pion,
$K_S^0$, proton and $\Lambda+\overline{\Lambda}$ $v_2$. The remaining
systematic uncertainties from detector artifacts are estimated by
comparing data taken with different field settings: 0.5 Tesla
(full-field) and 0.25 Tesla (half-field). The STAR experiment did not
collect half-field data during the 62.4 GeV data taking period so we
use the 200 GeV data to estimate the uncertainties in the 62.4 GeV
measurements. From these studies, we assign an uncertainty to $v_2$
for all particles of $\pm 0.0035$ (absolute).

The dominant systematic uncertainties in $v_2$ measurements arise from
correlations unrelated to the reaction plane (thought to be primarily
from correlations between particles coming from jets or resonance
decays or other correlations intrinsic to p+p collisions). When $v_2$
is measured using an event-plane analysis ($v_2\{EP\}$)~\cite{art},
these correlations can bias the experimental estimation of the
reaction plane (the event-plane angle) and change the apparent $v_2$
values (non-flow). A four-particle cumulant analysis of $v_2$
($v_2\{4\}$) is less sensitive to non-flow effects than a standard
analysis but yields larger statistical
uncertainties~\cite{Adler:2002pu}. Although $v_{2}\{4\}$ has been
shown to significantly reduce non-flow uncertainties, some sources of
uncertainty may remain: \textit{e.g.}  if $v_2$ fluctuates from
event-to-event $v_2\{4\}$ may yield values smaller than the mean
$v_2$~\cite{Miller:2003kd}. The magnitude of possible non-flow
correlations for unidentified charged hadrons is discussed in
Refs.~\cite{Adler:2002pu,Adams:2004bi}. Here we also discuss
variations of non-flow effects between different hadron types.

For $p_T<1$~GeV/c, a four-particle cumulant analysis is carried out
for pions and protons identified with greater than 98\% purity. To
study the hadron-type dependence of non-flow effects at intermediate
$p_T$, we analyze two samples of charged hadrons at $2.4 < p_T <
3.6$~GeV/c: one with $n\sigma_{\pi} > 0$, the other with $-5 <
n\sigma_{\pi} < -2.5$. Data from the 10\%--40\% centrality interval
are used. For $n\sigma_{\pi} > 0$, approximately 98\% of the
charged tracks are pions. For $-5 < n\sigma_{\pi} <
-2.5$, the sample contained approximately 75\% protons, 19\% kaons and
6\% pions. The ratio of the event-plane $v_2$ ($v_2$\{EP\})
to the cumulant $v_2$ ($v_2$\{4\}) for the pion sample and the proton
sample are listed in Table~\ref{nonf}. In the $p_T$ region below
1~GeV/c, proton $v_2$ does not appear to manifest any non-flow
correlations for either energy. For pions in this region, however,
non-flow correlations seem to account for ~10\% of the $v_2$ measured
with the event-plane analysis.

At intermediate $p_T$, $v_2\{EP\}/v_2\{4\}$ is greater than unity for
protons and pions. This shows that non-flow correlations increase the
observed $v_2\{EP\}$ for both protons and pions. At 62.4 GeV, the
increase is the same (within errors) for both particles. With the
larger 200 GeV data set however, we observe a larger non-flow
fraction for pions than protons: the pion $v_2\{EP\}/v_2\{4\} = 1.22
\pm 0.02$ and $v_2\{EP\}/v_2\{4\}$ for the proton sample $= 1.16 \pm
0.02$.  Pion $v_2\{EP\}$, therefore, appears to be more susceptible to
non-flow correlations than $v_2$ for particles in the proton sample.

\begin{table}[hbt]
\caption{The ratio $v_2$\{EP\}/$v_2$\{4\} ($v_2$ from a standard
  event-plane analysis over $v_2$ from a four-particle cumulant
  analysis) for the centrality interval 10\%--40\% in three $p_T$
  ranges (units for $p_T$ are GeV/c). The sample from $2.4<p_T<3.6$~GeV
  labeled as protons contains contamination from pions (6\%) and kaons
  (19\%). } \label{nonf}
\begin{tabular}{ccccc}
\toprule ~ & \multicolumn{2}{c}{62.4 GeV}
& \multicolumn{2}{c}{200 GeV}\\ 
\colrule 
	   ~~$p_T$~~ & pions & protons & pions & protons \\
\colrule 
$0.3-0.5$ & $1.09 \pm 0.01$ & $1.01 \pm 0.10$ & $1.10 \pm 0.01$ & $0.97 \pm 0.07$ \\ 
$0.5-0.7$ & $1.10 \pm 0.01$ & $0.98 \pm 0.08$ & $1.09 \pm 0.01$ & $0.99 \pm 0.05$ \\ 
$2.4-3.6$ & ~$1.08 \pm 0.04$~ & ~$1.11 \pm 0.05$ & ~$1.22 \pm 0.02$~ & ~$1.16 \pm 0.02$ \\ 
\botrule
\end{tabular}
\end{table}

\subsection{Baryon vs. Anti-baryon $v_2$}

To our knowledge, no prediction for a difference between baryon and
anti-baryon $v_2$ exists in the literature. Previous measurements at
RHIC of identified baryon $v_2$ reported no differences between
$\Lambda$ and $\overline{\Lambda}$ $v_2$ or between proton and
anti-proton $v_2$. Typically the particle and anti-particle samples
were combined. These measurements were made with smaller data samples
and at higher energies where the anti-baryon to baryon yield ratios
are much closer to unity.
Several scenarios can lead to a difference between anti-baryon and
baryon $v_2$ that is larger when the anti-baryon to baryon yield ratio
is smaller: (1) baryons may develop larger momentum-space anisotropies
through multiple rescattering as they are transported to mid-rapidity,
(2) if the initial spatial net-baryon density is anisotropic, flow
developing in a later stage could convert that spatial anisotropy to
an observable momentum-space anisotropy, and (3) annihilation of
anti-baryons in the medium can reduce the anti-baryon yield, with the
reduction larger in the more dense, in-plane direction than the
out-of-plane direction. We consider scenario (1) and (2) to be
distinct. In scenario (1), extra $v_2$ is built up while the baryons
are being transported to mid-rapidity, while in scenario (2) the $v_2$
is established through rescattering after the baryons are transported
to mid-rapidity.

\begin{figure}[hbtp]
\centering\mbox{
\includegraphics[width=0.5\textwidth]{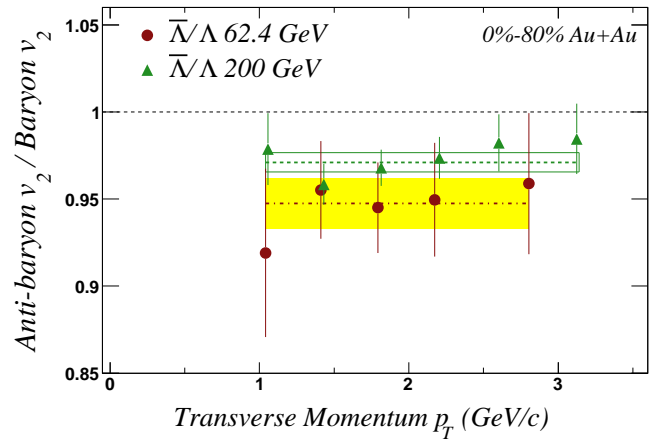}
}
\caption{ (color online). The ratio of $\overline{\Lambda}$ $v_2$ to
  $\Lambda$ $v_2$. The data are from minimum bias Au+Au collisions at
  $\sqrt{s_{_{NN}}}=$ 62.4 and 200 GeV. The bands show the average
  values of the ratios within the indicated $p_T$
  ranges. } \label{figbbarb}
\end{figure}

In Fig.~\ref{figbbarb} we show the ratio of $\overline{\Lambda}$ $v_2$
to $\Lambda$ $v_2$. The data are from minimum bias Au+Au collisions at
62.4 and 200 GeV. The bands on the figure represent the mean values of
the ratios which are respectively $0.948 \pm 0.014$ and $0.971 \pm
0.005$ for 62.4 and 200 GeV. In the measured range, the
$\overline{\Lambda}$ $v_2$ is systematically smaller than the
$\Lambda$ $v_2$ for both energies and within errors is approximately
$p_T$ independent (fitting the data with a straight line yields slopes
of $0.014 \pm 0.028$~(GeV/c)$^{-1}$ and $0.017 \pm
0.010$~(GeV/c)$^{-1}$ respectively for 62.4 and 200 GeV data). The
difference between $\overline{\Lambda}$ and $\Lambda$ $v_2$ is larger
at 62.4 GeV, where the $\overline{\Lambda}$ to $\Lambda$ yield ratio
is smaller. Taking into account the $\overline{\Lambda}/\Lambda$ yield
ratios (measured to be $0.532 \pm 0.014$ at 62.4 GeV and $0.77 \pm
0.05$ at 200 GeV/c~\cite{starbbarb}), we find that at 62.4 GeV the net
$\Lambda$ $v_2$ (the asymmetry of the quantity $\Lambda -
\overline{\Lambda}$) is 12\% $\pm$ 3\% larger than the $v_2$ of all
other $\Lambda$s or $\overline{\Lambda}$s. At 200 GeV it is 13\% $\pm$
4\% larger.
The larger $\Lambda$ $v_2$ is not anticipated from the RQMD hadronic
transport model~\cite{rqmd} where at mid-rapidity, the ratio of
anti-proton $v_2$ to proton $v_2$ is $1.148 \pm 0.084$ and the ratio
of $\overline{\Lambda}$ $v_2$ to $\Lambda$ $v_2$ is $1.142 \pm
0.123$. We note however that this model does not reproduce the overall
magnitude of $v_2$ at this energy either.

\begin{figure}[hbtp]
\centering\mbox{
\includegraphics[width=0.5\textwidth]{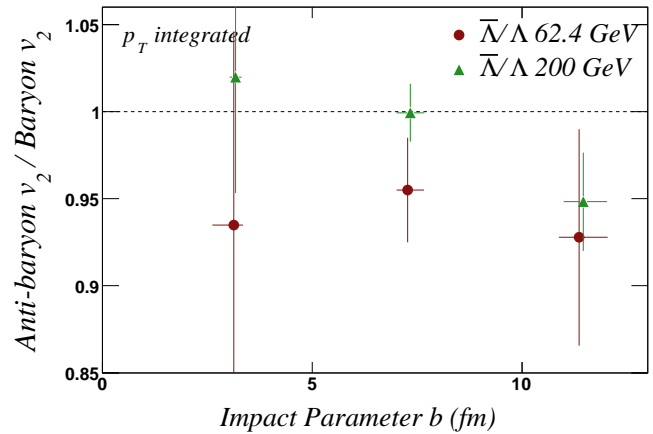}
}
\caption{ (color online). The $p_T$ integrated ratio of
  $\overline{\Lambda}$ $v_2$ to $\Lambda$ $v_2$ for three centrality
  intervals: 0\%--10\%, 10\%--40\%, and 40\%--80\%. The data are from
  Au+Au collisions at $\sqrt{s_{_{NN}}}=$ 62.4 and 200 GeV. } \label{figbbarbcent}
\end{figure}

In Fig.~\ref{figbbarbcent} we display the centrality dependence of the
$p_T$ integrated $\overline{\Lambda}$ $v_2$ to $\Lambda$ $v_2$
ratio. A Monte-Carlo Glauber model is used to convert the centrality
intervals defined by multiplicity into mean impact parameter
values. Given the errors we are unable to make a definitive statement
about a possible dependence of the ratio on centrality.

\subsection{Quark-number Scaling}

Models of hadron formation by coalescence or recombination of quarks
successfully reproduce many features of hadron production in the
intermediate $p_T$ region ($1.5<p_T<5$~GeV/c)~\cite{scalingv2, reco,
  CoalVoloshinv2}. These models find that at intermediate $p_T$, $v_2$
may follow a quark-number ($n_q$) scaling with $v_2(p_T/n_q)/n_q$ for
most hadrons falling approximately on one curve. In these models, this
universal curve represents the momentum-space anisotropy developed by
quarks prior to hadron formation. This scaling behavior was observed
in Au+Au collisions at 200 GeV~\cite{scalingv2}. Approximate quark
number scaling of $v_2$ also exhists in RQMD models where the scaling
is related to the additive quark hypothesis for hadronic
cross-sections~\cite{rqmd, Lu:2006qn}. The RQMD model, however,
under-predicts the value of $v_2$ by approximately a factor of
two. Pre-hadronic interactions are therefore thought necessary to
generate a $v_2$ as large as that observed at RHIC. If $v_2$ is
predominantly established in this pre-hadronic phase, the hadronic
cross-sections might not play a dominant role in establishing the
particle-type dependence of $v_2$.


\begin{figure*}[hbtp]
  \resizebox{0.45\textwidth}{!}{\includegraphics{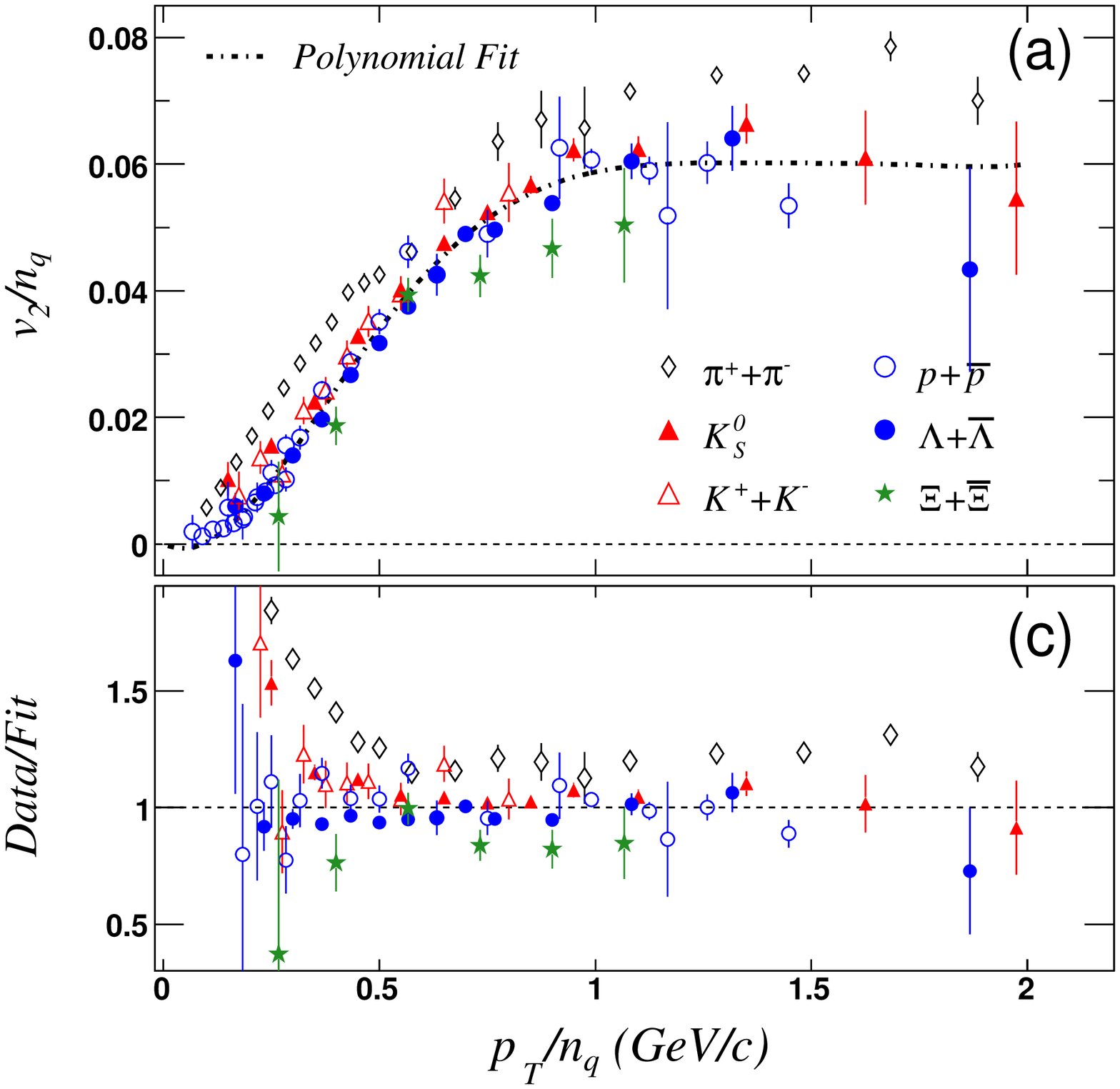}}
  \resizebox{0.45\textwidth}{!}{\includegraphics{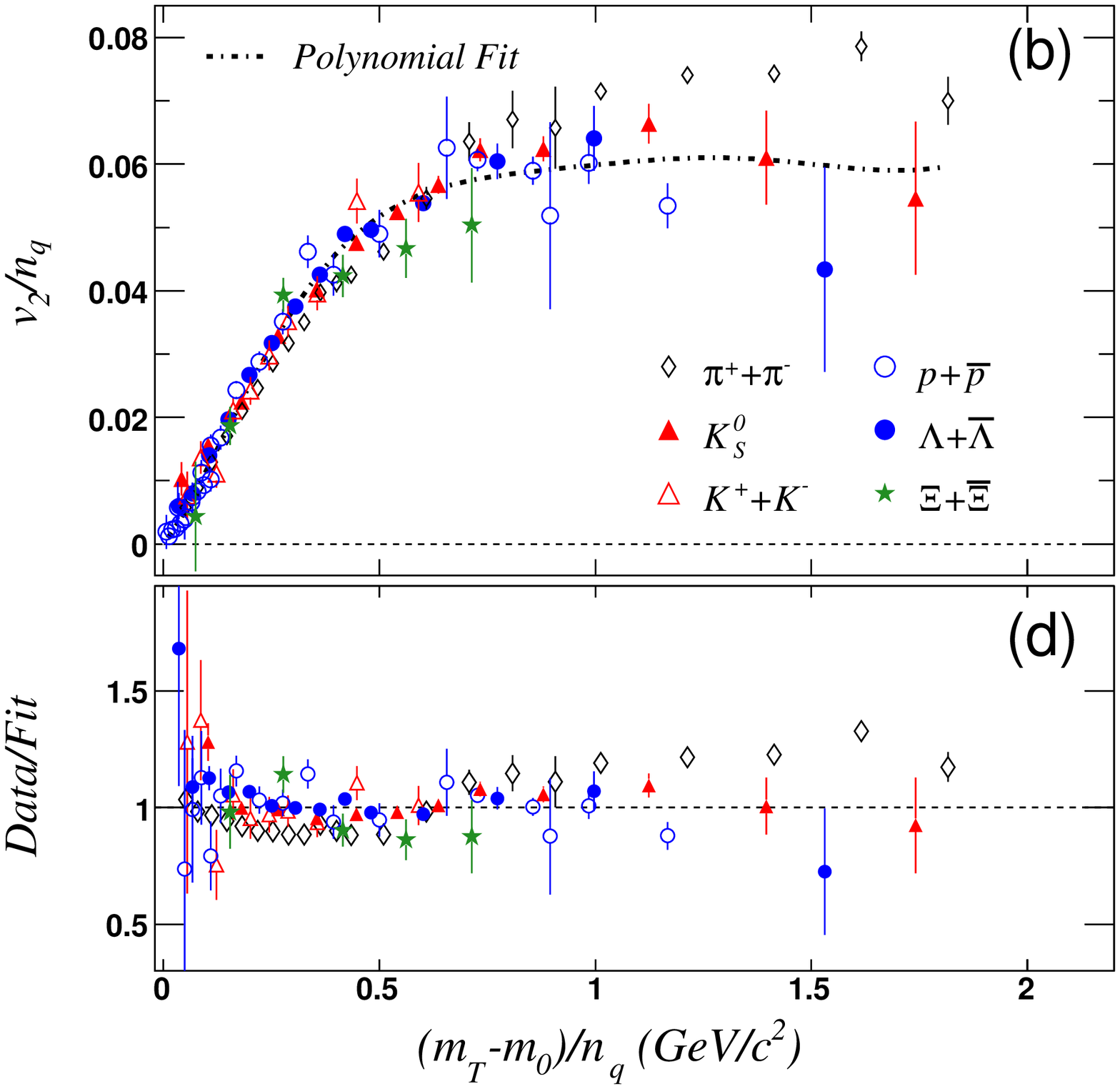}}
    \caption{(color online). Identified particle $v_2$ from minimum
      bias collisions at $\sqrt{s_{_{NN}}}=62.4$~GeV scaled by the
      number of valence quarks in the hadron ($n_q$) and plotted
      versus $p_T/n_q$ (a) and $(m_T-m_0)/n_q$ (b). In each case a
      polynomial curve is fit to all particles except pions. The ratio
      of $v_2/n_q$ to the fit function is shown in the bottom panels
      (c) and (d).}
    \label{fig3}
\end{figure*}

Fig.~\ref{fig3} shows $v_2$ scaled by the number of valence quarks in
the hadron ($n_q$) as a function of $p_T/n_q$ (a) and $(m_T-m_0)/n_q$
(b) for identified hadrons at $\sqrt{s_{_{NN}}}=62.4$~GeV. A
polynomial function has been fit to the scaled values of $v_2$ for all
particles except pions, which, for reasons discussed below, may
violate the scaling. To investigate the quality of agreement between
hadron species, the data from the top panel are scaled by the fitted
polynomial function and plotted in the bottom panels (c) and (d) of
Fig.~\ref{fig3}. In panel (c), for $p_T/n_q>0.6$~GeV/c, the scaled
$v_2$ of $K_S^0$, $K^{\pm}$, p+$\mathrm{\overline{p}}$ and
$\Lambda+\overline{\Lambda}$, lie on a single curve, within
errors. The 62.4 GeV data for these species are therefore consistent
with the scaling observed in 200 GeV collisions. The
$\Xi+\overline{\Xi}$ $v_2$ may lie below the curve but the current
errors do not permit a strong conclusion regarding deviations between
$\Xi+\overline{\Xi}$ and $p+\overline{p}$ or
$\Lambda+\overline{\Lambda}$ $v_2$. At $p_T/n_q<0.6$~GeV/c, the
scaling breaks down.

It was shown that for 200 GeV at $m_T-m_0<0.8$~GeV/c$^2$,
$v_2(m_T-m_0)$ is a linear function and independent of hadron
mass~\cite{sqm03}. In Fig.~\ref{fig3} panels (b) and (d) we combine
$m_T$ scaling and $n_q$ scaling so that a single curve can be used to
approximately describe $v_2$ throughout the measured range. This is
the same scaling as used in Ref.~\cite{Adare:2006ti} where the figures
are labelled $KE_T$ ($m_T-m_0$ is the transverse kinetic energy). This
combined scaling works because in the range where $v_2$ is a linear
function of $m_T-m_0$, dividing by $n_q$ does not alter the shape of
the curve. Once it is observed that $v_2$ for all particles follow the
same linear function for $m_T-m_0$, the scaling of $v_2(m_T-m_0)$ with
$n_q$ becomes trivial. At higher $p_T$, $v_2$ is only weakly dependent
on $p_T$ so that changing the axis variable from $p_T/n_q$ to
$(m_T-m_0)/n_q$ does not effect the scaling significantly.

Pion $v_2$ deviates significantly from the fit function in both panels
(a) and (b). The contribution of pions from resonance decays to the
observed pion $v_2$ may account for much of the deviation for $p_T <
1.5$~GeV/c~\cite{decayv2}. For $p_T>1.5$~GeV/c, non-flow correlations
discussed previously may contribute to the deviation. From the results
in Table~\ref{nonf}, we conclude that non-flow effects tend to be
larger for pions than protons. Particularly for the 200 GeV data,
removing non-flow contributions will increase the difference between
pion and proton $v_2$ and will improve the agreement between pion
$v_2/n_q$ and $v_2/n_q$ for the other measured particles. It has also
been suggested that constituent-quark-number scaling may be violated
for pions because the pion mass is much smaller than the masses of its
constituent-quarks. This implies a larger binding energy and a wider
wave-function for the pion. As a result, the approximation that
hadrons coalesce from constituent-quarks with nearly co-linear momenta
is broken~\cite{decayv2}.

\begin{figure*}[hbtp]
\centering\mbox{
\includegraphics[width=0.95\textwidth]{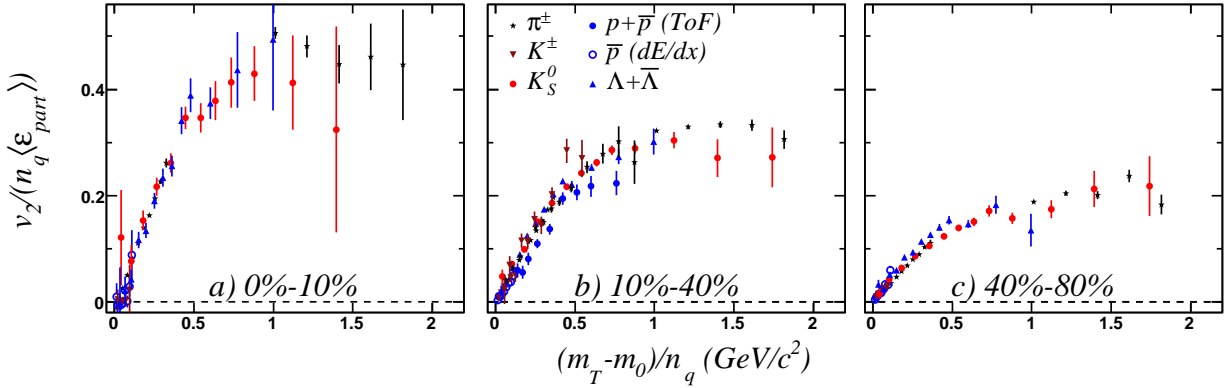}
}
\caption{(color online). $v_2/n_q$ scaled by the mean eccentricity of
  the initial overlap region versus $(m_T-m_0)/n_q$ for 0\%--10\%,
  10\%--40\%, and 40\%--80\% most central Au+Au collisions at
  $\sqrt{s_{NN}}$ = 62.4 GeV. \label{fig2b}}
\end{figure*}

Fig.~\ref{fig2b} shows $v_2/n_q$ versus $(m_T-m_0)/n_q$ for 0\%--10\%,
10\%--40\%, and 40\%--80\% most central Au+Au collisions at
$\sqrt{s_{NN}}$ = 62.4 GeV. $v_2/n_q$ for each centrality interval is
scaled by the mean eccentricity of the initial overlap region. The
eccentricity is calculated from the mean $x$ and $y$ positions of the
participating nucleons using a Monte-Carlo Glauber model. The
coordinate system is shifted and rotated so that $(0,0)$ is located at
the center-of-mass of the participants and the eccentricity is the
maximum possible. This is referred to as the participant eccentricity
($\varepsilon_{part}$). Since the true reaction plane is not know, our
$v_2$ measurements are sensitive to
$\varepsilon_{part}$~\cite{phobos}.  For the 0\%--10\%, 10\%--40\%,
and 40\%--80\% centrality intervals the
$\langle\varepsilon_{part}\rangle$ values respectively are 0.080,
0.247, and 0.547.

The $m_T-m_0$ and $n_q$ scalings shown for minimum bias data in
Fig.~\ref{fig3} are also valid within the specific centrality intervals
shown in Fig.~\ref{fig2b}. Early hydrodynamic calculations predicted
that $v_2$ should approximately scale with the initial spatial
eccentricity of the collision
system~\cite{Kolb:2000sd}. $v_2/\langle\varepsilon_{part}\rangle$ contradicts these
expectations and rises monotonically as the centrality changes from
peripheral to central. This indicates that central collisions are more
efficient at converting spatial anisotropy to momentum-space
anisotropy.

\subsection{Fourth Harmonic $v_{4}$}

\begin{figure}[hbtp]
\centering\mbox{
\includegraphics[width=0.5\textwidth]{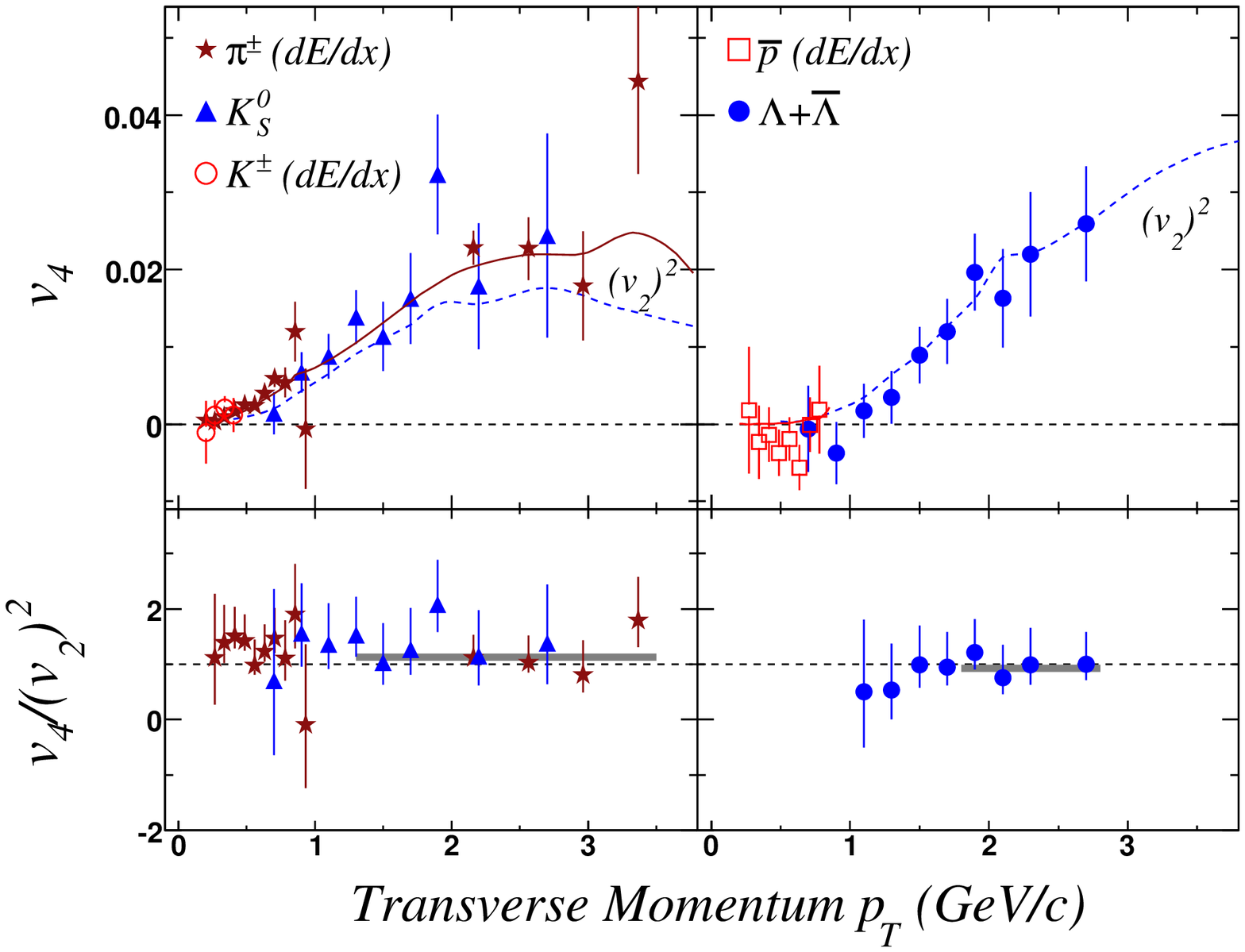}
}
  \caption{(color online). Top panels: minimum bias $v_4$ for pions,
    charged kaons, $K_S^0$, anti-protons and
    $\Lambda+\overline{\Lambda}$ at $\sqrt{s_{_{NN}}}=$ 62.4 GeV. In
    the left panel the solid (dashed) line shows the value for $v_2^2$
    for pions (kaons). In the right panel the dashed line is $v_2^2$
    for $\Lambda +\overline{\Lambda}$. Bottom panels: $v_4$ scaled by
    $v_2^2$ (points where $v_4$ and $v_2$ fluctuate around zero are
    not plotted). Grey bands correspond to the fit results described
    in the text and Table~\ref{v4tab}. The systematic errors on the
    $v_4/v_2^2$ ratio from non-flow are included in the error bars
    leading to asymmetric errors.
\label{fig5}}
\end{figure}

Higher order anisotropy parameters ($v_4$, $v_6$, etc.) may be
sensitive probes of hydrodynamic behavior and the initial conditions
of the collision system~\cite{Kolb:2003zi}. The authors of
Ref.~\cite{Borghini:2005kd} argue that values of the ratio $v_4/v_2^2$
larger than 0.5 indicate deviations from ideal fluid behavior. When
measured for identified particles, higher harmonics can also test
quark-number scaling~\cite{recov4}. $v_4$ and $v_6$ for charged
hadrons at 200 GeV are shown in Ref.~\cite{Adams:2003zg}. Identified
particle $v_4$ at 200 GeV is shown in Ref.~\cite{Adams:2004bi}. In
Fig.~\ref{fig5} (top panels) we plot pion, kaon, anti-proton and
$\Lambda+\overline{\Lambda}$ $v_4$ for $\sqrt{s_{_{NN}}}=62.4$ GeV,
where the standard event-plane analysis method has been used.  In the
bottom panels of Fig.~\ref{fig5} we show the ratio $v_4/v_2^2$ for
charged pions, neutral kaons, and hyperons. The uncertainty in
$v_4/v_2^2$ from possible non-flow leads to asymmetric errors. The
ratio $v_4/v_2^2$ is well above 0.5 even when errors are taken into
account.

In simple coalescence models~\cite{recov4}, the ratio $v_4/v_2^2$ for
hadrons is related to $v_4/v_2^2$ for quarks:
\begin{equation}
\left[ v_4/v_2^2\right]^{\mathrm{Meson}}_{2p_{T}} \approx 1/4 + (1/2)\left[
  v_4/v_2^2\right]^{\mathrm{Quark}}_{p_T}
\end{equation}
\begin{equation}
\left[ v_4/v_2^2\right]^{\mathrm{Baryon}}_{3p_{T}} \approx 1/3 + (1/3)\left[
  v_4/v_2^2\right]^{\mathrm{Quark}}_{p_T}
\end{equation}
where here $p_T$ is the quark $p_T$.  The $v_4/v_2^2$ for mesons can
also be related to $v_4/v_2^2$ for baryons:
\begin{equation}
\left[ v_4/v_2^2\right]^{\mathrm{Baryon}}_{3p_{T}} \approx 1/6 + (2/3)\left[
  v_4/v_2^2\right]^{\mathrm{Meson}}_{2p_T}
\end{equation}
Within this simple model, the large $v_4/v_2^2$ ratios presented here
indicate a large quark $v_4$.  At intermediate $p_T$, where
quark-scaling is thought to be valid, we use the equations above to
fit $v_4/v_2^2$ simultaneously for mesons and baryons, with
$v_4/v_2^2$ for quarks as a free parameter. The fit range is
$p_T>1.2$~GeV/c for mesons and $p_T>1.8$~GeV/c for baryons. A good
$\chi^2$ per degree-of-freedom (4.4/13) is found with quark
$v_4/v_2^2=1.93 \pm 0.29$. The grey bars in the bottom panels of
Fig.~\ref{fig5} show the corresponding $v_4/v_2^2$ values for mesons
and baryons. $\langle v_4/v_2^2 \rangle$ values for
$p_T/n_q>0.6$~GeV/c from data and the fit are listed in
Table~\ref{v4tab}. Since pion $v_2$ is known to deviate from the
simple scaling laws, we also performed the fit excluding the pion data
points (fit~II) which yielded a $v_4/v_2^2=2.18 \pm 0.40$ and $\chi^2$
per degree-of-freedom of 2.3/9. The small $\chi^2$ values for both
fits indicate that our data are consistent with quark-number scaling
where quark $v_4/v_2^2$ is approximately 2.

\begin{table}[hbt]
\caption{The ratio $v_4/v_2^2$ for $p_T/n_q > 0.6$~GeV/c from a
  combined fit and from data. Pion data points are used for fit~I and
  excluded for fit~II. The $\chi^2$ per degree-of-freedom is also
  shown on the bottom row.} \label{v4tab}
\begin{tabular}{cccc}
\toprule 
~~~~~ & data              & fit~I             & fit~II            \\ 
\colrule 
$\pi^{\pm}$
      &  $1.10 \pm 0.09$  &  $1.16 \pm 0.16$  & ~                 \\ 
$K^0_S$ 
      & ~$1.39 \pm 0.19$~ & ~$1.16 \pm 0.16$~ & ~$1.33 \pm 0.30$~ \\ 
~~$\Lambda + \overline{\Lambda}$~~ 
      &  $0.98 \pm 0.15$  &  $0.94 \pm 0.10$  &  $1.05 \pm 0.20$  \\ 
quark 
      &        ~          &  $1.93 \pm 0.29$  &  $2.18 \pm 0.40$  \\ 
\colrule
$\chi^2/dof$ 
      &        ~          &  $4.4/13$         &  $2.3/9$  \\ 

\botrule
\end{tabular}
\end{table}

\begin{figure}[hbtp]
\centering\mbox{
  \includegraphics[width=0.5\textwidth]{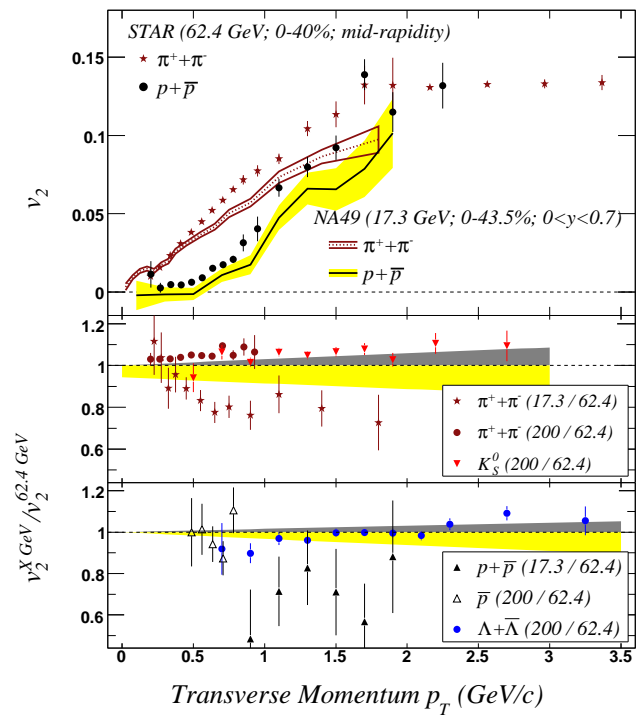}}
  \caption{(color online). Top panel: $v_2$ for pions and protons at
    $\sqrt{s_{_{NN}}}=$ 62.4 and 17.3 GeV. The 62.4 GeV data are from
    TOF and $dE/dx$ measurements combined. Middle and bottom panel:
    ratios of $v_2$ for $\pi^++\pi^-$, $K_S^0$, $p+\overline{p}$,
    $\Lambda+\overline{\Lambda}$ and at different center-of-mass
    energies scaled by the values at 62.4 GeV. The grey and yellow
    bands represent systematic uncertainties in the $v_2$ ratios
    arising from non-flow effects. The grey bands (above unity) are
    the uncertainties for the 200~GeV/62.4~GeV data and the yellow
    bands (below unity) are for the 17.3~GeV/62.4~GeV data.
\label{fig4}}
\end{figure}

\subsection{Collision Energy Dependence}

In Fig.~\ref{fig4} (top panel) we plot pion and proton $v_2$ from
$\sqrt{s_{_{NN}}}=62.4$ Au+Au and 17.3~GeV Pb+Pb
collisions~\cite{NA49}. In the bottom panels we show pion, $K_S^0$,
proton, and $\Lambda+\overline{\Lambda}$ data from 17.3 and/or 200 GeV
scaled by 62.4 GeV data. The 200 to 62.4 GeV ratios are taken using
$v_2$ data measured within the 0\%--80\% centrality interval. The TOF
$v_2$ measurements presented in this article allow us to show the 17.3
GeV to 62.4 GeV $v_2$ ratio to higher $p_T$ than the 200 GeV data
extends. In order to approximately match the centrality interval used
for the 17.3 GeV data, the 17.3 to 62.4 GeV ratios are taken using
respectively 0\%--43.5\% and 0\%--40\% centrality intervals. The STAR
data at 62.4 and 200 GeV are measured within the pseudo-rapidity
interval $|\eta|<1.0$ and the 17.3 GeV data are from the rapidity
interval $0<y<0.7$. These intervals represent similar $y/y_{beam}$
values. The same method is used to analyze the 200 and 62.4 GeV data.

Systematic errors from weak-decay feed-down and tracking errors will
mostly cancel when taking the ratio of $v_2$ at 200 and 62.4
GeV. Possible non-flow errors are larger at 200 GeV than at 62.4
GeV. In the lower panels of Fig.~\ref{fig4}, the shaded bands around
unity show the uncertainty in the energy dependence of the $v_2$ ratio
arising from possible changes in the magnitude of non-flow effects at
different energies. The portion of the band above unity applies to the
ratio of 200 and 62.4 GeV data while the portion below unity only
applies to the ratio of the 17.3 and 62.4 GeV data.

The $v_2$ data for pions and kaons at 62.4 GeV tends to be about 5\%
smaller than the 200 GeV data (although at $p_T>1$~GeV/c the
difference is within systematic uncertainties). The anti-proton data
at 62.4 and 200 GeV are consistent within errors. The data exclude a
proton $v_2$ variation between 62.4 and 200 GeV greater than
approximately 15\%. The $\Lambda+\overline{\Lambda}$ data show a
potentially interesting $p_T$-dependence: for $p_T<1.5$~GeV/c the 200
GeV $\Lambda+\overline{\Lambda}$ $v_2$ is systematically smaller than
the 62.4 GeV data while for $p_T>1.5$~GeV/c the 200 GeV
$\Lambda+\overline{\Lambda}$ $v_2$ data are consistent with or larger
than the 62.4 GeV data. Such a dependence can arise if the system in
200 GeV collisions develops a larger expansion velocity.

Appreciable differences are seen between the 17.3 GeV and 62.4 GeV
data. At $p_T>0.5$~GeV/c, for both pions and protons, the $v_2$ values
measured at 62.4 GeV are approximately 10\%--25\% larger than those
measured at 17.3~GeV~\cite{NA49, note2}. Although the
magnitude of $v_2$ is different at the lower energy, the systematics
of the particle-type dependencies are similar. In particular, pion
$v_2$ and proton $v_2$ cross over each other (or attain similar values)
at $p_T$ near $1.7$~GeV/c for $\sqrt{s_{_{NN}}}=$ 17.3, 62.4 and 200
GeV data. Due to the limited kinematic range covered by the 17.3 GeV
data, a quark-number dependence of $v_2$ at intermediate $p_T$ can
neither be confirmed nor excluded.

The increase in the magnitude of $v_2$ from 17.3 GeV to 62.4 GeV and
the similarity of 62.4 GeV $v_2$ to 200 GeV $v_2$ has been taken as a
possible indication for the onset of a limiting
behavior~\cite{ph62v2}. In a collisional picture, a saturation of
$v_2$ could indicate that for $\sqrt{s_{_{NN}}}$ at and above 62.4~GeV
the number of collisions the system constituents experience in a given
time scale can be considered large and that hydrodynamic equations can
therefore be applied. Hydrodynamic model calculations of $v_2$ depend
on the model initialization and the poorly understood freeze-out
assumptions~\cite{hydroPasi01, hydroShuryak01}. As such, rather than
comparing the predicted and measured values at one energy, the most
convincing way to demonstrate that a hydrodynamic limit has been
reached may be to observe the onset of limiting behavior with
$\sqrt{s_{_{NN}}}$. For this reason, $v_2$ measurements at a variety
of center-of-mass energies are of interest. Contrary to the large
differences reported in Ref.~\cite{ph62v2}, we find that when the 17.3
and 62.4 GeV $v_2(p_T)$ data are compared within similar
$|y|/y_{beam}$ ranges and when possible non-flow systematic
uncertainties are accounted for (the yellow bands in the bottom panel
of Fig.~\ref{fig4}), the differences between $v_2(p_T)$ within the
data sets may be as small as 10\%--15\%. As such, a large fraction of
the deviation between the SPS data and hydrodynamic models arises due
to the wide rapidity range covered by those measurements ($v_2$
approaches zero as beam rapidity is approached~\cite{Back:2004zg}),
increased $\langle p_T \rangle$ values at RHIC and the larger $v_2$
values predicted for the lower colliding energy by hydrodynamic
models.

\section{Conclusions}
\label{co}

We presented measurements of $v_2$ for pions, kaons, protons,
$\Lambda$, $\overline{\Lambda}$, $\Xi+\overline{\Xi}$, and
$\Omega+\overline{\Omega}$ from Au+Au collisions with
$\sqrt{s_{_{NN}}}=62.4$~GeV. We compared these measurements to similar
measurements at $\sqrt{s_{_{NN}}}=17.3$ and 200 GeV. The 62.4 GeV
pion, kaon, proton, and hyperon $v_2$ data are, within a few percent,
consistent with the equivalent data at 200 GeV. Within similar
$y/y_{beam}$ intervals and and after we account for systematic
uncertainties, we find that for a given identified particle species
the difference between 17.3 and 62.4 GeV $v_2$ data may be as small as
10\%--15\%. We find that $\Lambda$ $v_2$ is larger than
$\overline{\Lambda}$ $v_2$ at 62.4 and 200 GeV and that the difference
is larger at 62.4 GeV where the anti-baryon to baryon yield ratio is
smaller. At both energies our measurements are consistent with net
$\Lambda$ $v_2$ being approximately 10\%--15\% larger than
$\overline{\Lambda}$ and pair-produced $\Lambda$ $v_2$.

Our $v_2$ measurements at 62.4 GeV are consistent with the
quark-number scaling of $v_2$ first observed from Au+Au collisions at
$\sqrt{s_{_{NN}}}=200$~GeV. The 17.3 GeV data do not extend to high
enough $p_T$ to test quark-number scaling. We note, however, that the
$p_T$ where the $v_2$ values for mesons and baryons cross over each
other (or, in the case of 17.3 GeV data, become similar) is
approximately the same at all three center-of-mass energies. This
indicates that identified particle $v_2$ at 17.3 GeV may also be
consistent with quark-number scaling.

We also reported measurements of the higher harmonic term, $v_4$, for
pions, kaons, protons, and $\Lambda + \overline{\Lambda}$. These
measurements are also consistent with quark-number scaling laws
arising from coalescence or recombination models~\cite{recov4}. This
quark-number dependence may indicate that in ultra-relativistic
heavy-ion collisions collective motion is established amongst quarks
and gluons before hadrons are formed. This view is supported by the
large $v_2$ values measured for multi-strange baryons at
$\sqrt{s_{_{NN}}}=62.4$ and 200 GeV~\cite{msbv2}. Collisions involving
lighter nuclei and larger, deformed nuclei (U+U) will provide another
opportunity to study mass and quark number systematics for $v_2$.  The
possible approach to limiting values for $v_2$ (where the $p_T$ and
mass dependence at $p_T < 1$~GeV/c are consistent with hydrodynamic
models) along with the evidence presented here that the relevant
degrees of freedom in the early system may be sub-hadronic
(\textit{e.g. constituent quarks}) suggests that a strongly coupled
matter with sub-hadronic degrees of freedom may be created in
heavy-ion collisions at RHIC.

\section*{Acknowledgments}
We thank the RHIC Operations Group and RCF at BNL, and the
NERSC Center at LBNL for their support. This work was supported
in part by the Offices of NP and HEP within the U.S. DOE Office 
of Science; the U.S. NSF; the BMBF of Germany; CNRS/IN2P3, RA, RPL, and
EMN of France; EPSRC of the United Kingdom; FAPESP of Brazil;
the Russian Ministry of Science and Technology; the Ministry of
Education and the NNSFC of China; IRP and GA of the Czech Republic,
FOM of the Netherlands, DAE, DST, and CSIR of the Government
of India; Swiss NSF; the Polish State Committee for Scientific 
Research; SRDA of Slovakia, and the Korea Sci. \& Eng. Foundation.

\end{document}

%% file: authors.tex
%

\affiliation{Argonne National Laboratory, Argonne, Illinois 60439}
\affiliation{University of Birmingham, Birmingham, United Kingdom}
\affiliation{Brookhaven National Laboratory, Upton, New York 11973}
\affiliation{California Institute of Technology, Pasadena, California 91125}
\affiliation{University of California, Berkeley, California 94720}
\affiliation{University of California, Davis, California 95616}
\affiliation{University of California, Los Angeles, California 90095}
\affiliation{Carnegie Mellon University, Pittsburgh, Pennsylvania 15213}
\affiliation{University of Illinois, Chicago}
\affiliation{Creighton University, Omaha, Nebraska 68178}
\affiliation{Nuclear Physics Institute AS CR, 250 68 \v{R}e\v{z}/Prague, Czech Republic}
\affiliation{Laboratory for High Energy (JINR), Dubna, Russia}
\affiliation{Particle Physics Laboratory (JINR), Dubna, Russia}
\affiliation{University of Frankfurt, Frankfurt, Germany}
\affiliation{Institute of Physics, Bhubaneswar 751005, India}
\affiliation{Indian Institute of Technology, Mumbai, India}
\affiliation{Indiana University, Bloomington, Indiana 47408}
\affiliation{Institut de Recherches Subatomiques, Strasbourg, France}
\affiliation{University of Jammu, Jammu 180001, India}
\affiliation{Kent State University, Kent, Ohio 44242}
\affiliation{Institute of Modern Physics, Lanzhou, China}
\affiliation{Lawrence Berkeley National Laboratory, Berkeley, California 94720}
\affiliation{Massachusetts Institute of Technology, Cambridge, MA 02139-4307}
\affiliation{Max-Planck-Institut f\"ur Physik, Munich, Germany}
\affiliation{Michigan State University, East Lansing, Michigan 48824}
\affiliation{Moscow Engineering Physics Institute, Moscow Russia}
\affiliation{City College of New York, New York City, New York 10031}
\affiliation{NIKHEF and Utrecht University, Amsterdam, The Netherlands}
\affiliation{Ohio State University, Columbus, Ohio 43210}
\affiliation{Panjab University, Chandigarh 160014, India}
\affiliation{Pennsylvania State University, University Park, Pennsylvania 16802}
\affiliation{Institute of High Energy Physics, Protvino, Russia}
\affiliation{Purdue University, West Lafayette, Indiana 47907}
\affiliation{Pusan National University, Pusan, Republic of Korea}
\affiliation{University of Rajasthan, Jaipur 302004, India}
\affiliation{Rice University, Houston, Texas 77251}
\affiliation{Universidade de Sao Paulo, Sao Paulo, Brazil}
\affiliation{University of Science \& Technology of China, Hefei 230026, China}
\affiliation{Shanghai Institute of Applied Physics, Shanghai 201800, China}
\affiliation{SUBATECH, Nantes, France}
\affiliation{Texas A\&M University, College Station, Texas 77843}
\affiliation{University of Texas, Austin, Texas 78712}
\affiliation{Tsinghua University, Beijing 100084, China}
\affiliation{Valparaiso University, Valparaiso, Indiana 46383}
\affiliation{Variable Energy Cyclotron Centre, Kolkata 700064, India}
\affiliation{Warsaw University of Technology, Warsaw, Poland}
\affiliation{University of Washington, Seattle, Washington 98195}
\affiliation{Wayne State University, Detroit, Michigan 48201}
\affiliation{Institute of Particle Physics, CCNU (HZNU), Wuhan 430079, China}
\affiliation{Yale University, New Haven, Connecticut 06520}
\affiliation{University of Zagreb, Zagreb, HR-10002, Croatia}

\author{B.I.~Abelev}\affiliation{Yale University, New Haven, Connecticut 06520}
\author{M.M.~Aggarwal}\affiliation{Panjab University, Chandigarh 160014, India}
\author{Z.~Ahammed}\affiliation{Variable Energy Cyclotron Centre, Kolkata 700064, India}
\author{B.D.~Anderson}\affiliation{Kent State University, Kent, Ohio 44242}
\author{D.~Arkhipkin}\affiliation{Particle Physics Laboratory (JINR), Dubna, Russia}
\author{G.S.~Averichev}\affiliation{Laboratory for High Energy (JINR), Dubna, Russia}
\author{Y.~Bai}\affiliation{NIKHEF and Utrecht University, Amsterdam, The Netherlands}
\author{J.~Balewski}\affiliation{Indiana University, Bloomington, Indiana 47408}
\author{O.~Barannikova}\affiliation{University of Illinois, Chicago}
\author{L.S.~Barnby}\affiliation{University of Birmingham, Birmingham, United Kingdom}
% ...

\collaboration{STAR Collaboration}\noaffiliation

%% file: pid-v2-62.19.bbl
\begin{thebibliography}{99}
\bibitem{note} The reaction-plane angle is defined by the angle
  (relative to the laboratory frame) of the vector that connects the
  centers of the two colliding nuclei.
\bibitem{note2} We note that we are comparing $p+\overline{p}$
  $v_2(p_T)$ for data sets with very different $\overline{p}/p$ yield
  ratios and in this paper we have shown that the $\overline{\Lambda}$
  $v_2$ is smaller than the $\Lambda$ $v_2$.
\end{thebibliography}
